\documentstyle[preprint,aps]{revtex}
\begin{document}
\def\bra{\langle}
\def\ket{\rangle}
\draft
\preprint{\today}
\title{Varied Signature Splitting Phenomena in Odd Proton Nuclei}
\author{Yang Sun$^{(1,2)}$, Da Hsuan Feng$^{(1,2)}$ and Shuxian Wen$^{(3)}$}
\address {
$^{(1)}$Department of Physics and Atmospheric Science, Drexel University \\
Philadelphia, Pennsylvania 19104, USA \\
$^{(2)}$Physics Division, Oak Ridge National Laboratory \\
Oak Ridge, Tennessee 37831, USA \\
$^{(3)}$China Institute of Atomic Energy,
P.O.Box 275, Beijing, P.R.China }

\maketitle

\begin{abstract}
Varied signature splitting phenomena in odd proton rare earth nuclei are
investigated. Signature splitting as functions of
$K$ and $j$ in the angular momentum projection theory is explicitly shown
and compared with those of the particle rotor model. The observed deviations
from these rules are due to the band mixings. The recently measured
$^{169}$Ta high spin data are taken as a typical example where fruitful
information about signature effects can be extracted.
Six bands, two of which have not yet been observed,
were calculated and discussed in detail in this paper.
The experimentally unknown band head energies are given.
\end{abstract}

\newpage

Although there were many discussion in the past about
signature splitting \cite{Ko.83,SBG.87,RAJ.92,MJR.94},
they were mainly devoted
to odd neutron rare earth nuclei. One reason for this could be because
for such nuclei, bands are built up by nearly pure $i_{13\over 2}$
intruders. Therefore, their structures are rather simple.
For the odd proton nuclei, however, the Fermi level is surrounded by more
$j-$subshells, all of which have rather different characters and will interact
with each other, thus complicating the structure. It is also for this reason
that we believe that
additional information about signature splitting could arise
from such nuclei. On the other hand, modern detectors and techniques
allow one not only to analyze high spin data
in the yrast band but also
more side bands. That different bands within one nucleus
show distinguished behaviors is certainly a challenge and a crucial test
of the existing
nuclear theories. The recently reported measurement \cite{Ta169} of the odd
proton rare earth nucleus, $^{169}$Ta, which has provided
us with fruitful information
for our understanding of signature splitting, is an excellent example for
such a purpose.

Signature~\cite{BM.75} is a quantum number
specifically appearing in a
deformed intrinsic system. It is neither a universal quantum number
nor a concept in the conventional spherical shell model. It is
related to the invariance of a
system with quadrupole deformation under a rotation of $180^0$ around a
principal axis (e.g. the x-axis)
\begin{equation}
\hat R_x ~=~ e^{i\pi \hat J_x} .
\label{sigop}
\end{equation}
If the system is axially symmetric, only the rotation around any principal axes
other than the symmetry one can define the signature quantum number.
Thus, signature is a consequence of a deformed
system and corresponds to a ``deformation invariance" with respect to space
and time reflection \cite{Bo.76}. For the
even-even nuclei the signature operator $\hat R_x$ has two eigenvalues
$\pm 1$, while in the odd mass nuclei $\pm i$. By requiring
that the deformed core be
filled up from the bottom in the two-fold degenergate orbits with nucleons
having signature $\pm i$, the total signature of the ground-state band for
the even-even nucleus is $+1$. For an odd mass nucleus, depending on its total
spin, it can assume two different values. In fact, it is customary to assign
\begin{equation}
\alpha_I ~=~ {1\over 2} (-1)^{I-1/2}
\label{signo}
\end{equation}
as the signature quantum number for a state of spin $I$ of an odd mass nucleus.
Such a rotational band with a sequence of levels differing in spin by 1$\hbar$
is now divided into two branches, each consisting of levels differing in spin
by 2$\hbar$ and classified by the signature quantum number $\alpha_I = \pm
{1\over 2}$, respectively. Experimentally, one often observes an energy
splitting for the two branches.

The origin of the signature splitting in a rotational band can be understood
in the particle-rotor model. For the strong coupling limit,
the symmetrized wave function~\cite{BM.75} can be written as
\begin{equation}
\mid \phi^{IM}_{iK}>=\sqrt{{2I+1 \over{16\pi^2}}} \;
\Bigl\lbrace \; D^I_{MK} (\Omega ) \mid iK>+ \;
(-)^{I-K}D^I_{M-K} (\Omega ) \mid \overline{i K}> \; \Bigr\rbrace ,
\label{bmwavef}
\end{equation}
where $D^I_{MK}(\Omega )$ is the irreducible representation of the rotational
group, $\Omega$ the orientation of the rotor and $\mid \overline{i
K}>$ the time-reversed state of the particle state $\mid i K>$. The wave
function of eq.~(\ref{bmwavef}) has the factor $(-1)^{I-K}$ in its second
term. Taking the Coriolis coupling into account by the first order perturbation
theory, this $I$-dependent factor will appear in the energy spectrum and
contributes only to the $K = {1\over 2}$ band
\begin{equation}
E_{i, K={1\over 2}} ~ \sim ~ \Bigl\lbrace I(I+1) - a_i
(-1)^{I-{1\over 2}} (I+{1\over 2}) \Bigr\rbrace ,
\label{bmenergy}
\end{equation}
where $a_i$ is the so-called decoupling parameter \cite{BM.75}
and depends on the $j-$components which contribute to
the particle state $\mid i K={1\over 2}>$. From
the spectrum of eq.(\ref{bmenergy}), one sees that for a positive (negative)
decoupling parameter, the levels with odd (even) values of $I+{1\over 2}$ $(I =
{1\over 2}, {5\over 2}, {9\over 2}, ...  (I = {3\over 2},
{7\over 2}, {11\over 2}, ...  ))$ are shifted downwards, thus splitting one
band into two branches. The splitting amplitude and phase are respectively
determined
by the size and the sign of $a_i$.
This can explain the rather distorted bands for $K
= {1\over 2}$ in many odd mass nuclei within the context
of the particle-rotor model.

Of course, signature splitting is not necessarily confined to the
$K = {1\over 2}$ case. It appears in the $K = {1\over 2}$ band
in the particle-rotor model because
the Coriolis interaction couples bands which differ in $K$ quantum number by
$\pm 1$ (the $\Delta K = 1$ selection rule) and the Coriolis matrix has its
diagonal elements only between $K = {1\over 2}$ state and its time reversed
state in the first order perturbation theory.
However, as an observable
phenomenon, splitting of the energy into two branches
should also manifest itself in any nuclear
many-body theory.
To conform to the standard nomenclature, in this paper, we shall
keep using the name ``signature splitting".
In fact, it was shown \cite{HS.91b} that similar symmetrized
wave functions also exsit in the angular momentum projection theory
\cite{HI.79,HI.80}. Although, in the projection theory,
there is no explicit
decoupling parameter, the phenomenon of
signature splitting clearly shows up when an intrinsic single particle
state is projected onto states of good angular momentum.
One of the distinguished features of the angular momentum projection theory
is that the signature splitting can in fact persist
up to the $K = {5\over 2}$ band
\cite{HI.84,HS.92}, well beyond the $K = {1\over 2}$ band.
This is obviously important to explain
the data.

The angular momentum projection theory established in the late
seventies~\cite{HI.79,HI.80} has been proven to be a powerful model to
quantitatively account for many high spin phenomena~\cite{HS.91b,HS.92,HS.91a}.
Very recently, using this model, we have successfully explained
the anomalous crossing frequency in odd proton rare earth nuclei \cite{SFW.94}.
Since the model has already been discussed in detail elsewhere
\cite{HI.80,HS.91a,SFW.94}, we shall only outline
the major points of the theory which are relevant to the present discussions.

The ansatz for the angular momentum projected wave function is given by
\begin{equation}
| I M > ~=~
\sum_{\kappa} f_{\kappa} \hat P^I_{MK_\kappa}
| \varphi_{\kappa} > ,
\label{ansatz}
\end{equation}
where $\kappa$ labels the basis states. Acting on an intrinsic state $|
\varphi_{\kappa} >$, the projection operator
$\hat P^I_{MK}$ \cite{RS.80} generates states
of good angular momentum, thus restoring the necessary rotational symmetry
which was violated in the deformed mean field.
In the present work, we assume that the intrinsic states have axial symmetry.
Thus, the basis states $| \varphi_{\kappa} > $ must have $K$ as a good quantum
number.
For an odd proton system, the basis is spanned by the set
\begin{eqnarray}
\left\{
\;\; \alpha^\dagger_{p_l} |\phi >,
\;\;\; \alpha^\dagger_{n_i} \alpha^\dagger_{n_j} \alpha^\dagger_{p_l}
|\phi > \;\; \right\}
\label{baset}
\end{eqnarray}
The quasiparticle
vacuum is $|\phi >$ and $\{\alpha_m, \alpha^\dagger_m \}$ are the quasiparticle
annihilation and creation operators for this vacuum; the index $n_i$ (~$p_i$~)
runs over selected neutron (proton) quasiparticle states and $\kappa$ in
eq.~(\ref{ansatz}) runs over the configurations of eq.~(\ref{baset}). The
vacuum $|\phi >$ is obtained by diagonalizing a deformed Nilsson hamiltonian
\cite{An.78} followed by a BCS calculation. In the calculation, we have used
three major shells: i.e. N = 4, 5 and 6 (N = 3, 4 and 5) for neutrons (protons)
as the configuration space. The BCS blocking effect associated with the last
unpaired proton is approximately taken into account by allowing all the odd
number of protons to participate without blocking any individual level. Thus
the vacuum in this case is an average over the two neghboring even-even nuclei.
The size of the basis states
is determined by using energy windows of 1.5 MeV and 3 MeV for the 1- and 3-qp
states, respectively. Consequently, about 50 low-lying configurations are
constructed.

In this work, we have used the following hamiltonian
\cite{HI.80}
\begin{equation}
\hat H = \hat H_0 - {1 \over 2} \chi \sum_\mu \hat Q^\dagger_\mu
\hat Q^{}_\mu - G_M \hat P^\dagger \hat P - G_Q \sum_\mu \hat
P^\dagger_\mu\hat P^{}_\mu ,
\label{hamham}
\end{equation}
where $\hat H_0$ is the spherical single-particle shell model hamiltonian. The
second term is the quadrupole-quadrupole interaction and the last two terms are
the monopole and quadrupole pairing interactions respectively. The interaction
strengths are determined as follows: the quadrupole interaction strength $\chi$
is adjusted so that the known quadrupole deformation $\epsilon _2$ from the
Hartree-Fock-Bogoliubov self-consistent procedure \cite{Lamm.69} is obtained.
For example, for $^{169}$Ta it is 0.225; the monopole pairing strength
$G_M$ is adjusted to the known energy gap
\begin{equation}
G_M =  \left[ 20.12\mp 13.13 \frac{N-Z}{A}\right] \cdot A^{-1} ,
\label{GMONO}
\end{equation}
where the minus (plus) sign is for neutrons (protons). The quadrupole pairing
strength $G_Q$ is assumed to be proportional to $G_M$ and the proportional
constant is fixed to be 0.20 for all the bands calculated in the present work.

The weights $f_{\kappa}$ in eq.~(\ref{ansatz}) are determined by diagonalizaing
the hamiltonian $\hat H$ in the basis given by of eq.~(\ref{baset}). This will
lead to the eigenvalue equation (for a given spin $I$)
\begin{equation}
\sum_{\kappa '} ( H_{\kappa\kappa '} - E N_{\kappa\kappa '} )
f_{\kappa '} ~=~ 0 ,
\label{maha}
\end{equation}
with the hamiltonian and norm overlaps given by
\begin{eqnarray}
H_{\kappa \kappa'} &~=~& < \varphi_\kappa |\hat H \hat
P^I_{K_\kappa {K'}_{\kappa'}}|\varphi_{\kappa'} >,
 \nonumber \\
N_{\kappa\kappa'} &~=~& < \varphi_\kappa | \hat
 P^I_{K_\kappa {K'}_{\kappa'}} | \varphi_{\kappa'} > .
\label{norm}
\end{eqnarray}
The energies of each band are given by
the diagonal elements of eq.(\ref{norm})
\begin{equation}
E_\kappa (I) ~=~ \frac{ <\varphi_\kappa | \hat H \hat P^I_{KK} |
\varphi_\kappa >}{ < \varphi_\kappa | \hat P^I_{KK} |
\varphi_\kappa > } ~=~ \frac{ H_{\kappa\kappa}} {N_{\kappa\kappa} } .
\label{bandiag}
\end{equation}
A diagram in which $E_\kappa (I)$ of various bands are plotted against
the spin $I$ will be referred to \cite{HS.91a} as a band diagram.
Although the results obtained from diagonalizing the hamiltonian of
eq.(\ref{hamham}) can be compared with the data, globol behaviors
of the bands can already be understood by these diagonal elements,
as we shall see below.

We begin by showing the varied signature splittings
of the one quasiproton
bands (eq.(\ref{bandiag})) in Fig.1, where we have plotted
the energy differences
$E_I - E_{I-1}$ as a function of angular momentum. These bands are taken from
the present $^{169}$Ta calculation, thus reflecting
the realistic situations of the one
quasiproton bands in the rare earth region. In Fig.1a, which takes
the $1h_{11\over 2}$ $j-$subshell as an example, we show how the
splitting amplitude decreases with $K$ and increases with $I$. It is clearly
seen that the signature splitting is not restricted just to the $K={1\over 2}$
case. In fact, we see that the splitting
is large even for $K={5\over 2}$ where
a clear zig-zag at higher spins is observed.
Beyond $K\geq {7\over 2}$, the splitting is sufficiently
diminished.

Unlike Fig.1a, in Fig.1b, we have plotted six $K={1\over 2}$ bands from
different $j-$subshells. These bands are from $N=$ 4 and 5 proton major
shells, with some being close to the proton Fermi level and thus are
experimentally observed. Here the general feature is that the splitting
amplitude increases with increasing $j$ and the neighboring $j-$bands have
opposite splitting phases. This is consistent with the particle rotor model
(see eq.(\ref{bmenergy})). A particularly interesting case is the $2d_{3\over
2}$ case (${1\over 2}^+$[411] band).  For this band, the lower corners of the
curve due to the splitting are nearly zero, which implies a zero $\Delta I = 1$
transition energies. This implies also
that one should expect that there will be an
interesting energy degeneracy in the neighboring levels for this band, as we
will see below.

The band diagrams for $^{169}$Ta are plotted in Fig.2.
Although there are about 50
bands in the calculations, only several lowest lying ones
are plotted to illustrate the physics.
It is expected
that for $^{169}$Ta, the states ${1\over 2}^-$[541], ${7\over 2}^-$[523],
${9\over 2}^-$[514], ${1\over 2}^+$[411], ${5\over 2}^+$[402] and ${7\over
2}^+$[404] are close to the proton Fermi level and thus bands based on
these configurations are likely to be observed.
In Fig.2, each plotted band has
a definite $K$ quantum number. However, we stress that
$K$ will not be conserved due to the band mixing.
In fact, it is only an approximately conserved quantity,
depending on how strongly these bands are mixed by the diagonalization.
This is especially true for bands lying close in energy and with the
same symmetry. From
Fig.2 we will expect that at relatively low
spins, $K$ is approximately
a good quantum number because bands are separately distributed in this
spin region.
As the nucleus rotates more rapidly, level density in a given energy region
will be higher \cite{HS.91a} and the admixing of $K$ will be enhanced.
The amount of mixing can easily be analyzed from the wave functions.
Through the mixing, the bands near the yrast line
can display some of the properties of the higher lying bands.
In fact, the observed phenomenon signature inversion in the
odd-odd rare earth nuclei was explained
by the mechanism of band mixing \cite{HS.91b}.

In Fig.2a, we see that the
${9\over 2}^-$[514] band rises smoothly with a steeper slope (smaller moment
of inertia). Consequently, it crosses the 3-qp band at a lower spin (${31\over
2} \hbar$), which is consistent with the data \cite{Ta169}. Since this is a
high$-K$ band, one would not expect signature splitting to occur.
However, there is another band ${5\over
2}^-$[532] and its corresponding 3-qp band. The splitting for this
band is weak but distinct and will
admix with the ${9\over 2}^-$[514] band.
On the other hand, mixing of the ${1\over 2}^-$[541] band with its neighboring
bands is negligibly small because the former is originated
from different $j-$subshell.
Consequently, the experimentally observed small
splittings \cite{Ta169} in the ${9\over 2}^-$[514] band are clearly caused by
the mixing with ${5\over
2}^-$[532] band and therefore show the splitting phase of this band.

A rather different behavior is seen from the ${1\over 2}^-$[541] band: Due to a
strong signature effect, it is splitted into two branches and are separated
from each other by about 1 MeV.
Experimentally, the unfavored high-lying bands
are very weakly populated and therefore, in most cases,
only the favored branch with signature
$\alpha = {1\over 2}$ has been observed \cite{Ta169}. This is the common
observation for the ${1\over 2}^-$[541] band in the rare earth region. As
we have mentioned before, the zig-zag amplitude will
increase with spin.
This effect, as is shown in Fig.2a,
will bring the ${5\over 2} \hbar$ level to be lower in energy
than that of spin ${1\over 2} \hbar$, thus explaining
why the observed band head of ${1\over 2}^-$[541]
is usually ${5\over 2} \hbar$ and not ${1\over 2} \hbar$. Furthermore,
the band crossing of the ${1\over 2}^-$[541] with its 3-qp band
occurs at spin ${37\over 2} \hbar$, which again reproduces
the data \cite{Ta169}.
The ${7\over 2}^-$[523] band, which lies somewhat higher in energy than the
${1\over 2}^-$[541] favored branch,
has not been observed experimentally.

There are two types of interesting energy degeneracies in the positive parity
bands, as shown in Fig.2b. First, the two bands ${5\over 2}^+$[402] and
${7\over 2}^+$[404] are found to be nearly degenerate for the entire
band, thus making the experimental separation near impossible.
Although their interaction causes a slight
repulsion,
they nevertheless remain parallel and have nearly
identical moments of inertia. This will be shown clearly later on
in Fig.3. In this
situation, it is very difficult to distinguish them experimentally since they
have the same transition energies along their decay sequences.
In fact, only one of them was analyzed from
the data and was assigned as ${5\over 2}^+$[402] \cite{Ta169}. Thus, our
calculation clarifies as to why the most likely observed ${7\over
2}^+$[404] band in fact has escaped notice in the analysis of the data.
Clearly, this degeneracy is determined by the special
locations of the two bands
and thus should be isotope-dependent. It would be
interesting to
know which
one is favored in energy along the isotopic and the isotonic chains.

Second, as was pointed out in Fig.1b, the signature splitting in the special
configuration ${1\over 2}^+$[411] in $^{169}$Ta resulted in an
accidental
degeneracy between the two signature partners. Such a degeneracy implies
that the $\Delta I = 2$
transition energies along the two signature branches are identical. The reason
why only one signature branch in the ${1\over 2}^+$[411] band is
observed \cite{Ta169} is
not due to the large signature splitting as in the
${1\over 2}^-$[541] case, but by the accidental degeneracy. If this were true,
then the $\Delta I = 1$ transition energy between the two branches
(eg. ${5\over 2}^+ \rightarrow {3\over 2}^+$) should also
be identical to the $\Delta I = 2$ transition along the same branch
(eg. ${7\over 2}^+ \rightarrow {3\over 2}^+$). We anticipate that one could
ascertain these predictions by experimentally
distinguishing the
M1 and the E2 transitions
along the ${1\over 2}^+$[411] decay sequences. It should be noted
that in other odd proton rare earth nuclei, such accidental degeneracy may
not occur because the splitting is determined by the decoupling effect
which can be different from case to case.

All the three positive parity bands (${1\over 2}^+$[411], ${5\over 2}^+$[402]
and ${7\over 2}^+$[404]) cross their respective 3-qp bands roughly at spin
${29\over 2} \hbar$. The crossings change their
configurations from 1-qp to 3-qp. Indeed, the ${1\over 2}^+$[411] band
suddenly encounters a highly mixing region at spin ${29\over 2} \hbar$. This
may be why the measured band terminats at spin ${27\over 2}
\hbar$ \cite{Ta169}. The experimentally observed small splitting in the
${5\over 2}^+$[402] band at high spins \cite{Ta169}, from our analysis,
is due to
the mixing with ${3\over 2}^+$[411] band which
shows distinct zig-zag in its 3-qp configuration.

Finally, the entire calculated level scheme
for $^{169}$Ta is presented in Fig.3.
This scheme should be compared to the
experimental scheme given in Fig.2 of ref. \cite{Ta169}.
For completeness, the same diagram is given as Fig.4 of the present paper.
Three lowest lying positive and
negative parity bands are
shown here (see Fig.2 of the present work for their exact locations
before band mixing).
Two of them (${7\over 2}^+$[404] and ${7\over 2}^-$[523]) are our predictions.
Because of the missing linking transitions, the excitation energies of the band
heads cannot be determined experimentally. Our calculation suggests that
the $I = {5\over 2}$ of the band ${5\over 2}^+$[402]
is the lowest and is therefore set to be the reference level in Fig.3.
The excitation energies of the other band heads related to the reference level
are: 142 KeV ($I = {1\over 2}$ in ${1\over 2}^+$[411] band),
215 KeV ($I = {7\over 2}$ in ${7\over 2}^+$[404] band),
52 KeV ($I = {9\over 2}$ in ${9\over 2}^-$[514] band),
755 KeV ($I = {7\over 2}$ in ${7\over 2}^-$[523] band) and
807 KeV ($I = {5\over 2}$ in ${1\over 2}^-$[541] band).
The ${9\over 2}^-$[514] band has the lowest state at each spin and is therefore
the yrast band.
The nearly degenerate two signature branches of the band ${1\over 2}^+$[411]
is clearly seen. The two bands
${5\over 2}^+$[402] and ${7\over 2}^+$[404] (see Fig. 2b), due to their
interactions, are now shifted from each other in a parallel manner
by roughly 200 KeV. Yet, they maintain identical transition energies.
The two predicted bands, ${7\over 2}^+$[404] and ${7\over 2}^-$[523],
could be candidates for the experimentally observed but not yet assigned
band 5 in ref. \cite{Ta169}.
We should point out that at very high spins (around ${53\over 2}\hbar$),
one expects that the
5-qp states will cross the 3-qp states, thus lowering
the energy levels after the crossing.
Since such 5-qp configurations have not been included in
the present calculations, some of the computed high spin levels are found
to be too high in energy.

In conclusion, we have investigated the signature splitting phenomena in the
framework of the angular momentum projection theory.
The dependences of the splitting on the 3-component of the total angular
momentum $K$ and on the particle angular momentum $j$ are explicitly
given.
We have demonstrated
that in the projection theory, the signature splitting can persist beyond
the $K = {1\over 2}$ band and is also a diagonal effect for higher $K$
bands. These features are clearly different from the particle rotor model.
Furthermore, band mixing brings the behaviors of
higher lying band into the near yrast
bands, resulting in some observed deviations from the signature splitting
rules.

The $^{169}$Ta data are taken as an example to test our theory. Fruitful
information about signature splitting can be extracted from this nucleus.
We have pointed out that different explanations
as to why only
one signature branch is observed may work for the two $K = {1\over 2}$
bands: namely,
the accidental degeneracy of the two branches in the ${1\over 2}^+$[411] and
the usual large signature splitting in the ${1\over 2}^-$[541] band. Another
degeneracy which resulted in the identical transition energies after band
mixing, are found in ${5\over 2}^+$[402] and ${7\over 2}^+$[404] bands.
Finally, the entire calculated level scheme with six bands are given. We also
suggest the
band head excitations. Although the even-even nuclei have recently been
investigated in detail \cite{SE.94},
the present paper is the first application of the angular
momentum projection theory to the side bands in the odd-A system.

\acknowledgments
Useful discussions with
Mike Guidry and Lee L. Riedinger are acknowledged.
Yang Sun is most grateful to the College of Arts and Science of
Drexel University for the provision of a fellowship.
This work is partially supported by the United States National Science
Foundation.
Oak Ridge National Laboratory is managed by Martin Marietta Energy Systems,
Inc. for the U.S. Department of Energy under Contract No. DEAC05-84OR21400.

\baselineskip = 14pt
\bibliographystyle{unsrt}

\newpage
\begin{figure}
\caption{
Signature splitting as function of spin. a) Top: different $K$ bands of the
$1h_{11\over 2}$ subshell. b) Bottom: $K = {1\over 2}$ bands from different
$j-$subshells.}
\label{figure.1}
\end{figure}
\begin{figure}
\caption{
Band diagram for several lowest lying one-quasiproton and the corresponding
3-qp (one-quasiproton + neutron pair) bands in $^{169}$Ta. Note that the slope
of a curve $\omega = {dE\over dI}$ is the rotational frequency and its inverted
value measures the moment of inertia of the band. a) Left: negative parity
bands. b) Right: positive parity bands.
}
\label{figure.2}
\end{figure}
\begin{figure}
\caption{
Theoretical level scheme of $^{169}$Ta.
}
\label{figure.3}
\end{figure}
\begin{figure}
\caption{
Experimental level scheme of $^{169}$Ta. This is the same as the Fig.2 of
ref. \protect\cite{Ta169}.
}
\label{figure.4}
\end{figure}

\end{document}